\documentclass[%
 reprint,
 prl,
 floatfix,
 amsmath,amssymb,
 aps, widetext, superscriptaddress
]{revtex4-2}

\usepackage{xcolor}
\usepackage{hyperref}
\usepackage{graphicx}
\usepackage{dcolumn}
\usepackage{bm}
\usepackage{microtype}
\usepackage{physics}
\usepackage{xspace}
\usepackage{chemformula}
\usepackage{csquotes}

\newcommand{\changed}[1]{{#1}}

\renewcommand{\vec}[1]{\ensuremath{{\bm{#1}}}}

\allowdisplaybreaks[3]

\newcommand{\tuwien}{Institute for Theoretical Physics, Vienna University of Technology (TU Wien), Vienna A-1040, Austria}

\begin{document}
\title{Ultrafast Excitation Exchange in a Maxwell-Fish-Eye Lens}

\author{Oliver Diekmann}
\email{oliver.diekmann@tuwien.ac.at}
\author{Dmitry O. Krimer}
\author{Stefan Rotter}
\affiliation{\tuwien}

\date{\today}

\begin{abstract}
The strong coupling of quantum emitters to a cavity mode has been of paramount importance in the development of quantum optics. Recently, also the strong coupling to more than a single mode of an electromagnetic resonator has drawn considerable interest. We investigate how this multimode strong coupling regime can be harnessed to coherently control quantum systems. Specifically, we demonstrate that a Maxwell-Fish-Eye lens can be used to implement a pulsed excitation-exchange between two distant quantum emitters. This periodic exchange is mediated by single-photon pulses and can be extended to a photon-exchange between two atomic ensembles, for which the coupling strength is enhanced collectively.
\end{abstract}

\maketitle
\textit{Introduction.}---The coherent transfer of an excitation between different quantum systems is an essential process in quantum optics, with applications in quantum communication~\cite{gisin_quantum_2007}, quantum computation~\cite{van_meter_path_2016} and quantum networks~\cite{kimble_quantum_2008}. At the heart of many implementations lies the coupling to monochromatic light fields, permitting Rabi oscillations between the electromagnetic field and quantum emitters~\cite{jaynes_comparison_1963}.

Going beyond the paradigmatic Jaynes-Cummings model, the coupling of quantum emitters to more than a single mode has recently drawn considerable interest. A strong coupling to several modes is reached when the emitter-mode coupling strength becomes comparable to the free spectral range~\cite{meiser_superstrong_2006}. This multimode strong coupling regime (also known as the superstrong coupling regime) has been at the focus of recent theoretical studies~\cite{alber_photon_1992,meiser_superstrong_2006,krimer_route_2014, sinha_non-markovian_2020,mehta_theory_2022, hasebe_integrodifferential_2022} and experimental realizations, using photonic~\cite{bosman_multi-mode_2017,johnson_observation_2019,lechner_light-matter_2023,sundaresan_beyond_2015,puertas_martinez_tunable_2019,ferreira_collapse_2021,vrajitoarea_ultrastrong_2022, mehta_down-conversion_2023} or phononic modes~\cite{han_multimode_2016,moores_cavity_2018}.

The profound impact of single mode effects on quantum technology raises the question, whether mechanisms of similar scope can also be engineered making use of many modes in a cavity. In contrast to the single-mode case, the dynamics in multimode cavities are in general more complex as they are dictated by a complex interference of all involved modes, whose individual properties are determined by the geometry of the underlying resonator. This in turn suggests that a suitable choice for the cavity geometry is the key to tame the dynamics and harness it for the coherent control of quantum systems. 

\begin{figure}[b]  
    \centering
    \includegraphics[]{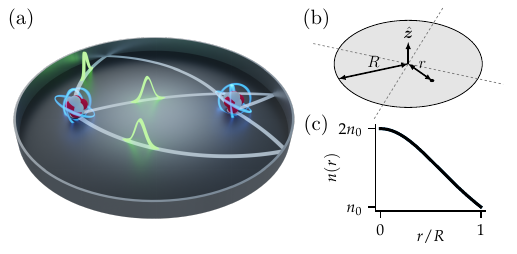}
    \caption{(a) 2D Maxwell-Fish-Eye lens with two opposing quantum emitters. Within the lens, all rays emerging from one atom form circular arcs due to the radial refractive index gradient, and converge at the second atom~\cite{leonhardt_perfect_2009}. All light paths that connect the two atoms with one reflection at the bounding mirror have the same optical length. (b) Illustration of relevant coordinates and parameters in the lens. (c) Radial refractive index gradient within the lens [cf. Eq.~\eqref{eq:MWFL_n}].}
    \label{fig:1}
\end{figure}

In this work, we demonstrate that with a Maxwell-Fish-Eye (MFE) lens~\cite{maxwell_solution_1854,tai_maxwell_1958} as a resonator, it is possible to coherently transfer an excitation between two distant quantum emitters when making use of their strong coupling to multiple modes. 
As a gradient index lens, the MFE lens comprises a radially changing refractive index and comes with the remarkable property that any point in the lens has a corresponding focal point mirrored at the lens' center. From the viewpoint of geometrical optics, light rays propagating in the lens form circular arcs, and all rays emitted from one point converge at the antipodal point [see Fig.~\ref{fig:1}(a)]. The presence of a continuum of pairs of focal points provides decisive advantages when considering distributed emitters, or emitters that cannot be considered point-like. Due to their extraordinary properties, MFE lenses have been studied, e.g., for imaging~\cite{leonhardt_perfect_2009, leonhardt_perfect_2010, leonhardt_perfect_2011, tyc_perfect_2011, merlin_maxwells_2011}, coherent perfect absorption~\cite{yin_multiple_2020} as well as for radiation emission~\cite{gevorkian_discrete_2020}; also generalizations of the MFE-concept have been proposed~\cite{demkov_internal_1971,tyc_absolute_2011,eskandari_elliptical_2019, davtyan_maxwell_2021}. In the context of quantum optics, the MFE lens has recently come into focus for the creation of entanglement by the dipole-dipole interactions it may mediate~\cite{perczel_quantum_2018}. In contrast, we consider the MFE lens in a regime, where the \changed{light field in the} cavity exhibits strong memory effects.

Within the multimode strong coupling regime, it has been shown both theoretically and experimentally that a {\it single} excited atom may emit its excitation into the electromagnetic field and (partially) reabsorb the corresponding single-photon pulse when it returns to the atom~\cite{krimer_route_2014,sanchez_munoz_resolution_2018,
ferreira_collapse_2021,hasebe_integrodifferential_2022}. Hence, when such a pulse is emitted by an excited atom in a MFE lens, one may expect the pulse to be refocused at the opposite point. In the following, we consider a second atom placed at this opposite point, and study the dynamics of the two quantum emitters. Taking into account the quantized electromagnetic field inside the cavity, we demonstrate explicitly the emission of a single-photon pulse, its propagation through the cavity and its refocusing at the second atom. In the most straightforward implementation, the pulse does not get fully absorbed by the second atom, which we attribute to a mismatch in the temporal pulse shape~\cite{stobinska_perfect_2009, aljunid_excitation_2013, trautmann_time-reversal-symmetric_2015, leong_time-resolved_2016}. By spectrally engineering the mode-emitter coupling, we can increase the efficiency of this exchange to a value near unity, thus implementing a pulsed passive swap operation~\cite{bechler_passive_2018} that relies on the interaction with many modes. On that basis, we then show that the dynamics of two emitters is readily translated to two atomic ensembles, where the coupling strength is collectively enhanced~\cite{johnson_observation_2019}. Our results demonstrate how concepts from classical photonics can be translated to the quantum regime when emitters couple strongly to many modes.

\textit{The model.}---We model the above system using the Glauber-Lewenstein quantization~\cite{glauber_quantum_1991, perczel_quantum_2018} where the dynamics is described by the Hamiltonian
\begin{equation}
H=\hbar\omega_\mathrm{a}\sum_{i}\ketbra{\mathrm{e}_i}{\mathrm{e}_i}+\hbar\sum_{lm}\omega_{l}a^{\dagger}_{lm}a^{\phantom{\dagger}}_{lm}-\sum_{i}\vec{d}_i\vec{E}(\vec{r}_i)\,,
\end{equation}
comprising the two-level atoms, the quantized field and the coupling between them, respectively. Here, $\omega_\mathrm{a}$ is the atoms' transition frequency and the excited (ground) state of the $i$th emitter is denoted by $\ket{\mathrm{e}_i}$ ($\ket{\mathrm{g}_i}$). Since we will extend the discussion from two atoms to two atomic ensembles, we keep the number of atoms general at this point. The frequencies of the cavity modes are denoted by $\omega_{l}$ and the modes' creation and annihilation operators obey the usual commutation relations. We note that, while the above interaction Hamiltonian is conventionally used, there has been an ongoing debate regarding its gauge invariance for multimode systems~\cite{arwas_properties_2023, stokes_implications_2022}. The key results of our present work are, however, left unchanged by this issue.

The radially varying refractive index in the MFE lens is given by
\begin{equation}
    n(\vec{r})=\frac{2n_0}{1+(r/R)^2},
    \label{eq:MWFL_n}
\end{equation}
where $r$ is the in-plane distance from the lens' center and $R$ is the radius of the mirror, which forms the circular lens' boundary \changed{(see Fig.~\ref{fig:1})}. Without loss of generality, we set the minimal refractive index $n_0=1$ throughout the manuscript \changed{and assume the transverse thickness $b$ of the lens to be far below the lens' radius $R$}. \changed{The relevant transverse eigenmodes $\vec{f}_{lm}(\vec{r})$ in this refractive index profile are hence polarized transversal to the plane and} known analytically~\changed{\cite{perczel_quantum_2018}}.
The modes are indexed by $l=1,2,3,...$ and angular index $m=-(l-1),-(l-3),...,(l-1)$ and the modes' eigenfrequencies are $\omega_{l}=c\sqrt{l(l+1)}/(Rn_0)$, i.e., the $l$th mode is $l$-fold degenerate.
The emitters couple to the modes via the transverse electric field,
\begin{align}
\vec{E}(\vec{r}) = \vec{E}^{+}(\vec{r})+\vec{E}^{-}(\vec{r})\,,\quad \vec{E}^{-}(\vec{r})=\left[\vec{E}^{+}(\vec{r})\right]^\dagger\,,\notag\\
\vec{E}^{+}(\vec{r}) = i\sum_{lm}\sqrt{\frac{\hbar\omega_{l}}{2\varepsilon_0}}a_{lm}\vec{f}_{lm}(\vec{r})\,,\label{eq:e+}
\end{align}
and the dipole moment, $\vec{d}_i={d}(\sigma_i^{\dagger}+\sigma_i)\hat{\vec{z}},~\sigma_i=\ketbra{\mathrm{g}_i}{\mathrm{e}_i}$, of the atoms. We assume a Gaussian cutoff in frequency for the atomic dipole moments $d^2=d^2_0\exp[-(\omega-\omega_\mathrm{a})^2/(2\omega_\mathrm{c}^2)]$, with cutoff frequency $\omega_\mathrm{c}$~\cite{berman_spontaneous_2010,krimer_route_2014, sanchez_munoz_resolution_2018}. Further, we employ the rotating wave approximation, i.e., we neglect non-excitation preserving terms $a\sigma$, $a^\dagger\sigma^\dagger$. Note that this approximation is not necessarily guaranteed to be compatible with the multimode dynamics~\cite{hegerfeldt_causality_1994,ben-benjamin_causality_2020}, \changed{since ultrastrong coupling effects may eventually become relevant~\cite{sanchez_munoz_resolution_2018,falci_ultrastrong_2019, forn-diaz_ultrastrong_2019,giannelli_optimized_2022, vrajitoarea_ultrastrong_2022, bosman_multi-mode_2017}} \changed{(an explicit justification for the validity of the approximation for our work will thus be provided at the end of the paper).} The mode-independent coupling prefactors are summarized in the constant $g = d_0\sqrt{\omega_\mathrm{a}^3/(\varepsilon_0 b c^2\hbar)}$. While $g$ parametrizes the coupling strength between modes and atoms, it does not represent the couplings' actual values, which are reduced by mode specific prefactors [see Eq.~\eqref{eq:e+}]. \changed{We further note that for coupling strengths far below the free spectral range, Rabi oscillations between two emitters can be recovered in accordance with~\cite{perczel_quantum_2018}.}

To study a pulsed excitation transfer, we consider a single excitation only, i.e., we can represent the wavefunction as $\ket{\psi(t)} = \sum_i c_i(t)\ket{0\dots01_i0\dots0}_\mathrm{a}\ket{0...0}_\mathrm{ph}+\sum_{lm}c^\mathrm{ph}_{lm}(t)\ket{0\dots0}_\mathrm{a}\ket{0...01_{lm}0...0}_\mathrm{ph}$, comprising atomic and photonic degrees of freedom, and the dynamics can be calculated by standard tools~\cite{supplement}. Using this wavefunction, the observable electric field intensity \changed{$\expval{\vec{E}^-(\vec{r})\vec{E}^+(\vec{r})}$ is directly accessible}
and subsequently serves as an observable for the field in the cavity~\cite{haizhen_field_2004,purdy_manifestation_2003}.

\begin{figure}
    \centering
    \includegraphics{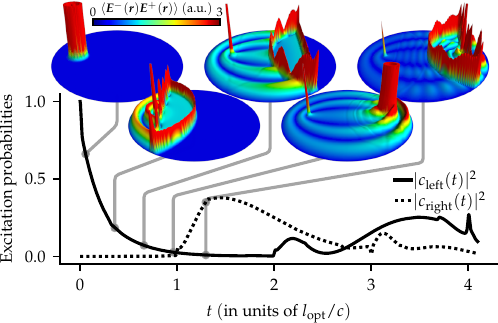}
    \caption{Atom and multimode dynamics in a Maxwell-Fish-Eye lens. The excited state populations of the two emitters are shown as a function of time. The insets display the intensity expectation value of the quantized electric field {$\expval{\vec{E}^-(\vec{r})\vec{E}^+(\vec{r})}$ } for five different times, as indicated by the links to the curves. Note that peaks of exceeding height were clipped in the insets. A video illustrating the dynamics is provided in~\cite{supplement}.}
    \label{fig:mfl}
\end{figure}

\textit{Single-photon pulses.}---We consider a MFE lens of radius $R=3\lambda_\mathrm{a}=6\pi/\omega_\mathrm{a}$ with two emitters placed opposite to one another at $r_\mathrm{{a}}=0.6R$ (see Fig.~\ref{fig:1}). The coupling strength is $g=0.5\omega_\mathrm{a}$ and the cutoff frequency is set to $\omega_\mathrm{c}=2\omega_\mathrm{a}$, i.e., the cutoff function is flat around the atomic transition frequency~\cite{krimer_route_2014}, leading to about $10^5$ modes participating in the dynamics. Figure~\ref{fig:mfl} shows the  resulting excitation probabilities of the two atoms as a function of time when the left emitter is initially excited. Over some time, the emitter transfers the excitation to the cavity and returns to its ground state. Subsequently, the excitation probability of the right hand emitter starts to increase when the time $t=l_\mathrm{opt}/c=R\pi n_0/c$ is reached. This time precisely corresponds to the propagation duration of light from one emitter to the other one, including one reflection at the cavity boundary~\cite{supplement}.

Here, since we have full access to the quantized photonic degrees of freedom, we can directly access the dynamics inside the cavity between the initial emission and the following absorption. The insets in Fig.~\ref{fig:mfl} show the expectation value of the intensity at different times. During the initial emission, a radially symmetric pulse emerges. This single-photon pulse propagates and is reflected wherever it strikes the cavity boundary. Notably, due to the refractive index gradient, the wavefront deforms during propagation. Since the emitters' positions constitute a pair of focal points, the pulse eventually refocuses on the second emitter, and is partly absorbed. \changed{Note that the spot size of the focused pulse may be smaller than the atomic wavelength $\lambda_\mathrm{a}$ since the emitter also interacts with high frequency modes and acts as a drain for the light field~\cite{ma_towards_2018,pichler_random_2019}.} \changed{Before the absorption by the right emitter, both emitters are close to their ground state. The absorption thus witnesses significant memory effects in the light field which would not be compatible with a Born-Markov approximation~\cite{perczel_quantum_2018}.}

While the pulse emitted by the left atom is successfully refocused on the right atom, the excitation is not fully transferred there. This can be understood by considering that the absorption process is the time-reverse of the emission process, which has been studied in detail in free-space setups~\cite{stobinska_perfect_2009, aljunid_excitation_2013, trautmann_time-reversal-symmetric_2015, leong_time-resolved_2016}. Since the emission from the left atom is of an exponential-like shape, the resulting pulse has an extended decaying tail. For perfect absorption, however, the time-reversed pulse, i.e., an increasing tail followed by the wavefront peak, would need to arrive at the right emitter. For the dynamics in Fig.~\ref{fig:mfl}, the mismatch in the temporal pulse shape results in an overlay between the absorption and the reemission dynamics for the right atom, i.e., when the excitation probability of the right atom reaches its peak, a new pulse is already propagating towards the left atom (see rightmost inset in Fig.~\ref{fig:mfl}). This process leads to a fast degradation of the pulsed excitation exchange between the two atoms.
\begin{figure}
    \centering
    \includegraphics{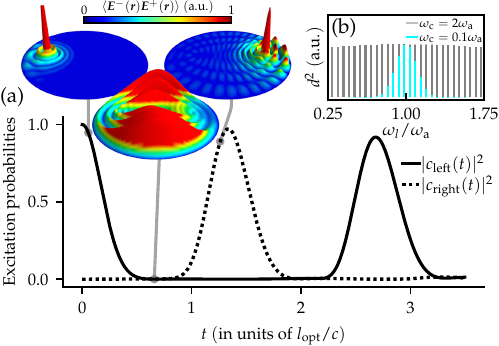}
    \caption{Coherent excitation exchange in the Maxwell-Fish-Eye lens. (a) The panel is structured in analogy to Fig.~\ref{fig:mfl}. All parameters but the cutoff frequency $\omega_\mathrm{c}=0.1\omega_\mathrm{a}$ are left unchanged as compared to Fig.~\ref{fig:mfl}. Due to the narrower range of modes participating in the dynamics, a high exchange efficiency is achieved. (b) Illustration of the spectral engineering ($\omega_\mathrm{c}=0.1\omega_\mathrm{a}$) in comparison with the cutoff used in Fig.~\ref{fig:mfl}. A video illustrating the dynamics is provided in~\cite{supplement}.}
    \label{fig:mfl_coherent}
\end{figure}

\textit{Coherent exchange by spectral engineering.}---Previous work suggests that a viable path to overcome the restrictions of the pulse shape is to artificially modify the pulse emission and absorption~\cite{cirac_quantum_1997, stobinska_perfect_2009, aljunid_excitation_2013, trautmann_time-reversal-symmetric_2015, leong_time-resolved_2016}. In the following, we show that in our setup we can achieve a nearly coherent transfer of the excitation by spectrally engineering the coupling strength between the emitters and the modes of the resonator. Specifically, we modify the couplings such that the atoms couple to a reduced number of about $200$ modes (amounting to about $15$ frequencies with degeneracies) close to the resonance frequency $\omega_\mathrm{a}$. Such an engineered coupling strength may be reached, e.g., by off-resonantly driving Raman-transitions with higher lying states~\cite{law_arbitrary_1996,cirac_quantum_1997}. Alternatively, by employing Bragg mirrors~\cite{bitton_two-dimensional_2018, sletten_resolving_2019, hutner_nanofiber-based_2020}, the coupling may effectively be reduced to modes lying in the bandgap, thus realizing the coupling to the restricted range of modes.

We parametrize the engineered coupling strength by reducing the cutoff frequency to $\omega_\mathrm{c}=0.1\omega_\mathrm{a}$ [see Fig.~\ref{fig:mfl_coherent}(b)], which results in a symmetrically distributed Gaussian coupling. In Fig.~\ref{fig:mfl_coherent}(a), we consider the corresponding dynamics for the left emitter initially excited. We now find that after the propagation, the excitation is almost completely reabsorbed by the second emitter. 
This greatly enhanced absorption efficiency can be understood when considering the shape of the emitted pulse shown in the intensity distributions in Fig.~\ref{fig:mfl_coherent}(a). Due to the spectral engineering, the emitted pulse is now symmetric around its peak, and therefore time-reversal symmetric. Consequently, the pulse is absorbed as efficiently by the right emitter as it was emitted by the left emitter. Without the spectral engineering, the coupling constants showed a pronounced asymmetry, since, first, the relevant modes' frequencies were bounded by zero from below, but extended far beyond the atomic transition frequency, and, second, the coupling constants obeyed their natural frequency dependence $\sim\sqrt{\omega_l}$ [see Eq.~\eqref{eq:e+}]. 
\changed{Even with the spectral engineering, we see that the excitation probability at the right atom peaks slightly below unity, which can be shown to be due to an imperfect time-reversal symmetry of the emitted pulse. This remaining infidelity can be reduced by further optimization of the coupling, cutoff and geometrical parameters together with the details of the coupling constants' frequency dependence~\cite{supplement}. }

Note that while the spectral engineering reduced the number of modes, the multimode nature of the dynamics remains apparent by the pulsed emission and absorption. Further, the emitter-cavity system remains strongly non-Markovian as in between the emission and absorption both emitters reach their ground state and the complete excitation is reversibly stored in the electromagnetic field.

Since both atoms emit pulses that can be absorbed efficiently, an autonomous periodic excitation swapping is established. Remarkably, since any point is a focal point in the MFE lens, the pulses between the emitters do not suffer from dispersion during propagation. The mechanism studied in Fig.~\ref{fig:mfl_coherent} further works universally also for MFE lenses with larger diameters.

\begin{figure}
    \centering
    \includegraphics{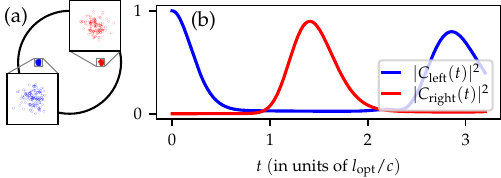}
    \caption{Collective excitation exchange between two atomic ensembles in a Maxwell-Fish-Eye lens. (a) Spatial distribution of the emitters in the lens. The two ensembles opposing each other consist of $N=100$ atoms each. The atoms are normal-distributed with a standard deviation of $\sigma=0.02R$ around the two centers at $r_\mathrm{{a}}=0.6R$ at opposite positions in the cavity. (b) Collective excitation probabilities of the ensembles $|C_\mathrm{left (right)}|^2=\sum_i|c_{i,\mathrm{left (right)}}|^2$ as a function of time. Initially, the excitation is distributed among the left ensemble as a symmetric Dicke state. The individual coupling strength of each atom of the ensemble is $g=0.05\omega_\mathrm{a}$ while the other parameters are analogous to Fig.~\ref{fig:mfl_coherent}.}
    \label{fig:ManyEmitters}
\end{figure}
\textit{Collective excitation exchange.}---Experimentally, it may be hard to reach the multimode strong coupling regime with single emitters. It is thus worthwhile to point out that this regime has recently been reached through collective enhancement of the coupling between an ensemble of atoms and a cavity~\cite{johnson_observation_2019}.

To showcase, that the pulsed excitation exchange by means of dielectric cavities also applies to collective multimode strong coupling, we replace each emitter by a cloud of atoms. As for the single emitters, we place the centers of the ensembles opposite to one another in the MFE lens. Within each ensemble, the atoms are normal-distributed around the ensemble center [see Fig.~\ref{fig:ManyEmitters}(a)]. 

In Fig.~\ref{fig:ManyEmitters}, we consider $N=100$ atoms per ensemble and reduce the coupling strength by a factor of $\sqrt{100}$ as compared to Fig.~\ref{fig:mfl_coherent}. When the left ensemble is initially prepared in a symmetric Dicke state, $\ket{\psi(0)} = \sum_{i=1}^{N}\ket{0...01_i0...0}_\mathrm{a,left}\ket{0...0}_\mathrm{a, right}\ket{0...0}_\mathrm{ph}/\sqrt{N}$, we can clearly observe a pulsed excitation swapping between the two ensembles comparable to Fig.~\ref{fig:mfl_coherent}. This shows that the exchange mechanism is robust when a distributed ensemble is considered. Even when the distribution of the atoms is further broadened, the pulsed dynamics remains intact, although with reduced efficiency. On the other hand the dynamics in Fig.~\ref{fig:ManyEmitters}(b) can be brought arbitrarily close to the one in Fig.~\ref{fig:mfl_coherent}(a) by narrowing the ensemble distribution.

\begin{figure}
    \centering
    \includegraphics{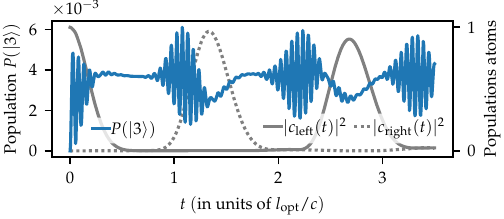}
    \caption{\changed{Population of the three-excitation subspace $P(\ket{3})$ (blue) when counter-rotating terms are taken into account. The dynamics of the atomic populations (grey) is shown in the background for comparison to Fig.~\ref{fig:mfl_coherent}. All parameters are identical to Fig.~\ref{fig:mfl_coherent}.}}
    \label{fig:fig-05}
\end{figure}

\changed{\textit{Rotating wave approximation.}---Since the relevant mode frequencies may deviate significantly from the atomic transition frequency $\omega_\mathrm{a}$, it is not a-priori clear whether counter-rotating terms can be neglected in the interaction between atoms and modes. In the following, we therefore explicitly verify the applicability of the rotating wave approximation to the setup of Fig.~\ref{fig:mfl_coherent}, which represents the principal result of our paper. In first order, the counter-rotating terms couple the single-excitation subspace to the three-excitation subspace. Hence, we take into account in our simulation all states that contain at most three excitations and monitor the cumulative population $P(\ket{3})$ of the states with exactly three excitations as a function of time. Figure~\ref{fig:fig-05} shows that $P(\ket{3})$  remains around $5\times 10^{-3}$, i.e., the dynamics is largely unaffected by counter-rotating terms.  
}

\textit{Conclusions.}---We demonstrated that a Maxwell-Fish-Eye lens can be used to transfer an excitation between distant quantum emitters in the multimode strong coupling regime. Our analysis shows that this exchange relies on the emission of single-photon pulses and can be highly efficient with the application of judicious spectral engineering of the atom-cavity coupling. In the absence of quantum emitters, the Maxwell-Fish-Eye lens is well understood in classical photonics; our study shows, that in the multimode strong coupling regime this intuition also applies to the resonator's quantum properties, thus providing a promising platform for multimode quantum physics. Interesting extensions to more than two emitters could be implemented with generalizations of the traditional MFE lens~\cite{demkov_internal_1971,tyc_absolute_2011,eskandari_elliptical_2019, davtyan_maxwell_2021}.\\

\begin{acknowledgments}
The authors would like to thank Joachim Sch\"oberl and Javier del Pino for helpful discussions.
OD gratefully acknowledges financial support by the Studienstiftung des Deutschen Volkes and the Austrian Science Fund (FWF) Grant No. P32300. The computational results presented have been
achieved using the Vienna Scientific Cluster (VSC). The following numerical packages have been employed in this work:~\cite{wieczorek_shtools_2018, johansson_qutip_2013}.
\end{acknowledgments}
\bibliography{Amfel.bbl}
\end{document}


\renewcommand{\thefigure}{S\arabic{figure}}
\setcounter{figure}{0}

\renewcommand{\theequation}{S\arabic{equation}}
\setcounter{equation}{0}
\title{Supplementary Information:\\Ultrafast Excitation Exchange in a Maxwell-Fish-Eye Lens}

\author{Oliver Diekmann}
\email{oliver.diekmann@tuwien.ac.at}
\author{Dmitry O. Krimer}
\author{Stefan Rotter}
\affiliation{\tuwien}

\date{\today}

\maketitle
\section{Single excitation dynamics}
In this supplementary section we give details about the calculation of the dynamics in the multimode cavity. As described in the main text, we employ a rotating-wave-approximated Hamiltonian~\cite{glauber_quantum_1991,perczel_quantum_2018},
\begin{align}
    H=&\hbar\omega_\mathrm{a}\sum_{k}\ketbra{\mathrm{e}_k}{\mathrm{e}_k}+\hbar\sum_l\omega_la^{\dagger}_la^{\phantom{\dagger}}_l\notag\\&\quad-i\hbar \sum_{k,l}\left[g_l^{(k)}\sigma_k^+a_l-g_l^{(k)*}\sigma_k^-a_{l}^{\dagger}\right]\,,
\end{align}
where we index all modes $lm$ by a single index $l$, for simplicity. We further introduced the atom-mode coupling constants $g_l^{(k)}=g\sqrt{{\omega_lc^2b}/(2{\omega^3_\mathrm{a}})}\mathcal{F}(\omega_l)\vec{f}_l(\Vec{r}_k)\hat{\vec{z}}$, with the Gaussian cutoff $\mathcal{F}^2(\omega)= {\exp[-(\omega-\omega_\mathrm{a})^2/(2\omega_\mathrm{c}^2)]}$. Since the Hamiltonian is excitation preserving and we only consider a single excitation in the system, we can write the wavefunction as 
\begin{align}
    \ket{\psi(t)} &= \sum_k c_k(t)\ket{0\dots01_k0\dots0}_\mathrm{a}\ket{0...0}_\mathrm{ph}\notag\\&\quad+\sum_lc^\mathrm{ph}_{l}(t)\ket{0\dots0}_\mathrm{a}\ket{0...01_l0...0}_\mathrm{ph}\notag\\
    &\equiv\sum_k c_k(t)\ket{k_\mathrm{a}}+\sum_lc^\mathrm{ph}_{l}(t)\ket{l_\mathrm{ph}}
\end{align}
Using the Schr\"odinger equation we can derive the dynamical equation for the coefficients in the wavefunction,
\begin{align}
&\dot{c}_{k}(t)
=\frac{1}{i\hbar}\left(\mel{k_\mathrm{a}}{H}{k_\mathrm{a}}{c}_{k}(t)+\sum_lc^\mathrm{ph}_l(t)\mel{k_\mathrm{a}}{H}{l_\mathrm{ph}}\right),\\
&\dot{c}^\mathrm{ph}_l(t)=\frac{1}{i\hbar}\left(c^\mathrm{ph}_l(t)\mel{l_\mathrm{ph}}{H}{l_\mathrm{ph}}+\sum_kc_k(t)\mel{l_\mathrm{ph}}{H}{k_\mathrm{a}}\right).
\end{align}
For $N$ atoms and $L$ relevant modes of the electromagnetic field, this system of differential equations can be rephrased in matrix-vector notation,
\begin{align}
i\hbar\dot{\vec{c}}=\vec{\mathcal{H}}\vec{c},
\end{align}
with the coefficient vector of dimension $N+L$,
\begin{equation}
    \vec{c}=\left(
    c_1(t),\dots,
    c_N(t), c_1^\mathrm{ph}(t),
    \dots,
    c_L^\mathrm{ph}(t)
    \right)\,,
\end{equation}

and the effective Hamiltonian

\begin{align}
    \vec{\mathcal{H}}=
    \left(
    \begin{matrix}
    \hbar\omega_\mathrm{a}&0&\dots&0&\alpha_{1,1}&\hdots&\hdots&\alpha_{1,L}\\
    0&\ddots&\ddots&\vdots&\vdots&\ddots&\ddots&\vdots\\
    \vdots&\ddots&\ddots&0&\vdots&\ddots&\ddots&\vdots\\
    0&\hdots&0&\hbar\omega_\mathrm{a}&\alpha_{N,1}&\hdots&\hdots&\alpha_{N,L}\\
    \alpha^{*}_{1,1}&\dots&\dots&\alpha^{*}_{N,1}&\hbar\omega_1&0&\hdots&0\\
    \vdots&\ddots&\ddots&\vdots&0&\ddots&\ddots&\vdots\\
    \vdots&\ddots&\ddots&\vdots&\vdots&\ddots&\ddots&0\\
    \alpha^*_{1,L}&\dots&\dots&\alpha^*_{N,L}&0&\hdots&0&\hbar\omega_L\\
    \end{matrix}
    \right)\,,
\end{align}
where we introduced $\alpha_{k,l} = -i\hbar g_l^{(k)}$.
The corresponding dynamics can be solved by standard numerical tools even for a comparatively large number of atoms and modes.
\section{Optical path length in the Maxwell-Fish-Eye lens}
\begin{figure}
    \centering
    \includegraphics{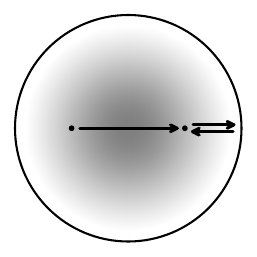}
    \caption{Ray trajectory used for the calculation of the optical path between any two focal points in a MFE lens.}
    \label{fig:sfig-01}
\end{figure}

In the Maxwell-Fish-Eye (MFE) lens, all paths between a pair of focal points with a single reflection share the same optical length,  which is clear from its construction via stereographic projection~\cite{leonhardt_perfect_2009}. 
To determine the optical path between any two emitters we can therefore select the simplest path that runs from an arbitrary point via the origin to the boundary, and then to the corresponding focal point (see Fig.~\ref{fig:sfig-01}). Correspondingly, the optical length can be calculated as 
\begin{equation}
    l_\mathrm{opt} = \int\limits_{-R}^R dr~ n(r)=\int\limits_{-R}^R dr \frac{2n_0}{1+r^2/R^2}=\pi n_0 R.
\end{equation}
It must be stressed that this optical length holds for any pair of focal points irrespective of their location in the lens. 
\changed{
\section{Parameter optimization}

To achieve high fidelity in the coherent excitation exchange within the MFE lens, the parameters of the emitter-cavity system have to be tuned in such a way that the excitation is transferred via time-reversal symmetric pulses. For this, it is most important to carefully adjust the coupling strength $g$ and the cutoff frequency $\omega_\mathrm{c}$. For Fig.~3 in the main text, we adjusted this pair of parameters by hand, which implies a remaining infidelity of $2.8\%$ for the first transfer of the excitation from the left atom to the right one. Here, we want to discuss methods to mitigate this remaining infidelity. 
\begin{figure}
    \centering
    \includegraphics{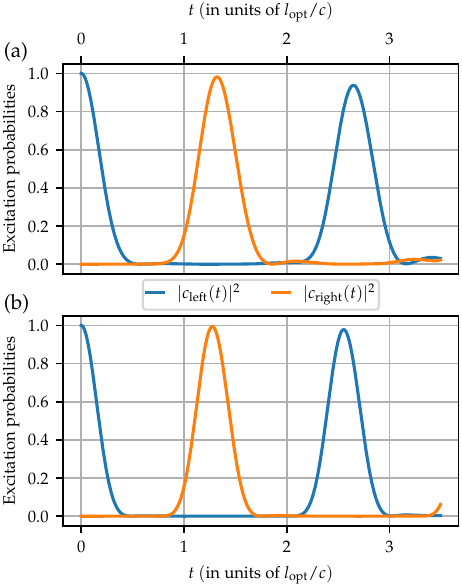}
    \caption{\changed{Excitation transfer after numerical minimization of the transfer infidelity. (a) The coupling constant $g$ and cutoff frequency $\omega_\mathrm{c}$ are taken as degrees of freedom in the optimization. After minimization, the infidelity is reduced to $1.9\%$. The corresponding setup parameters are given by $g=0.522\omega_\mathrm{a}$, $\omega_\mathrm{c}=0.0916\omega_\mathrm{a}$, $R=3\lambda_\mathrm{a}$ (not optimized) and $r_\mathrm{a}=0.6\lambda_\mathrm{a}$ (not optimized). (b) The frequency dependence of the coupling constants is slightly changed (see text) and $g, \omega_\mathrm{c}$, $R$ and $r_\mathrm{a}$ are taken as degrees of freedom in the optimization, starting from the parameters of Fig.~3 in the main text as an initial guess. Thus, the exchange infidelity is reduced to $0.66\%$. As optimized parameters we find $g=0.589\omega_\mathrm{a}$, $\omega_\mathrm{c}=0.0960\omega_\mathrm{a}$, $R=3.00\lambda_\mathrm{a}$ and $r_\mathrm{a}=0.586R$}.}
    \label{fig:s2}
\end{figure}

As a first step, we numerically minimize the infidelity with the coupling $g$ and the cutoff frequency $\omega_\mathrm{c}$ as parameters. In Fig.~\ref{fig:s2}(a), the remaining infidelity of the first exchange is already reduced to $1.9\%$. We note that, in principle, the Gaussian modulation of the coupling constants in our manuscript is still overlaid with the natural frequency dependence, $\sqrt{\omega_l}$, stemming from the electric field operator (cf. Eq.~(3) in the main text). This additional dependence causes a slight asymmetry in the distribution of the coupling constants around the atomic transition frequency. Eliminating the $\sqrt{\omega_l}$ factor, and adding the diameter of the lens together with the precise radial placement of the emitters to the degrees of freedom for optimization, we can arrive at infidelities as low as $0.66\%$ in Fig.~\ref{fig:s2}(b). This demonstrates that the infidelity in the excitation exchange can largely be reduced by fine-tuning of system parameters.
}

\bibliography{Bmfel_Supplement.bbl}